\documentclass[a4paper,12pt]{article}

\usepackage{amsmath}
\usepackage{amssymb}
\usepackage[dvips]{graphicx}
\usepackage[dvips]{psfrag}

\makeatletter
\@addtoreset{equation}{section}
\renewcommand{\theequation}{\thesection.\@arabic\c@equation}
\makeatother

\makeatletter
\renewcommand\appendix{\par%\newpage
  \setcounter{section}{0}%
  \setcounter{subsection}{0}%
  \gdef\thesection{Appendix \@Alph\c@section }
  \renewcommand{\theequation}
  {\Alph{section}.\arabic{equation}}
}
\makeatother

\begin{document}

\begin{titlepage}

\vspace*{-15mm}   
\baselineskip 10pt   
\begin{flushright}   
\begin{tabular}{r}    
{\tt KUNS-2342}\\ 
\end{tabular}   
\end{flushright}   
\baselineskip 24pt   
\vglue 10mm   

\begin{center}
{\Large\bf
 Entropy Balance Equation of\\ Spacetime Thermodynamics in f(R) Gravity
}

\vspace{8mm}   

\baselineskip 18pt   

\renewcommand{\thefootnote}{\fnsymbol{footnote}}

Yuki~Yokokura$^a$\footnote[4]{yokokura@gauge.scphys.kyoto-u.ac.jp}

\renewcommand{\thefootnote}{\arabic{footnote}}
 
\vspace{5mm}   

{\it  
 $^a$ Department of Physics, Kyoto University, 
 Kitashirakawa, Kyoto 606-8502, Japan 
 
}
  
\vspace{10mm}   

\end{center}

\begin{abstract}
We study spacetime thermodynamics for non-equilibrium processes. 
We first generalize the formulation of spacetime thermodynamics by using an observer outside the horizon.
Then we construct the entropy balance equation of spacetime thermodynamics for non-equilibrium processes in f(R) gravity. 
The coefficients of the expansion and shear terms are equal to the viscosities of the black hole membrane paradigm, 
and a new entropy production term appears.

\end{abstract}

\baselineskip 18pt   

\end{titlepage}

\newpage
%%%%%%%%%%%%%%%%%%%%%%%%%%%%%%%%%%%%%%%%%%%%%%%%%%%%%%%%%%%%%%%%%%%%%%%%%%%%%%%%%%%%%%%%%%%%
\section{Introduction}
Black hole solutions originally came from the Einstein equation. 
The four laws of the mechanics were analogous to those of thermodynamics\cite{Bardeen:1973gs}. 
With the discovery of the quantum Hawking radiation\cite{Hawking:1974sw}, it became clear that the analogy is an identity, and that black holes are thermodynamic objects. 
Among those, the result of Gibbons and Hawking left a mystery\cite{Gibbons:1976ue}. 
In the paper, black hole's entropy$~S=\frac{1}{4}A$ is derived from the free energy for the canonical system of a black hole, 
by using WKB approximation in Euclidean field theory. 
Then the finite statistical entropy results from a single classical black hole configuration. 
This may indicate that a solution of the Einstein equation corresponds to a thermodynamic state.

One might think that the above thermodynamic nature of spacetime is not restricted in the black hole spacetime. 
This speculation was investigated by Jacobson, 
and he concluded that, even non-black-hole spacetime has some thermodynamic property in the sense that 
the Einstein equation plays a role as ``the equation of state"\cite{Jacobson:1995ab}. 
He considered a part of any spacetime as a thermodynamic system 
by using the fact that a uniformly accelerating observer at any point in arbitrary spacetime has his own horizon (see the next section). 
He assumed the Uuruh effect\cite{Unruh:1976db}, the entropy area law, local equilibrium, quasi-statistical process, and that the all energy is the heat $(\delta E = \delta 'Q)$. 
From the Raychaudhuri equation and the Clausius definition of entropy $(T\delta S = \delta Q)$, he derived the Einstein equation. 
In this sense, the Einstein equation can be regarded as the equation of state. 

However, Jacobson's observer is strange. 
His observer is inside the horizon (inside the system) and measures energy flow into the system (see the next section). 
This is contrary to the spirit of thermodynamics because thermodynamic quantities are conventionally measured by an external observer. 
Therefore it is difficult to apply Jacobson's observer to a system of a black hole, 
but this fact is undesirable because Jacobson's idea should be general enough to be applicable to black hole thermodynamics. 
Note that Padmanabhan generalized Jacobson's idea to more general theories of gravity 
by using observers outside the horizon and Wald's entropy, 
though the formulation is different from Jacobson's and useful only to quasi-static processes\cite{Padmanabhan:2009vy}. 

We apply an outside observer to a dynamical spacetime, use the Raychaudhuri equation, 
generalize the formulation of spacetime thermodynamic system to the extent that we can consider non-equilibrium processes from viewpoint of the outside observer. 
Then we construct the entropy balance equation of spacetime thermodynamics for non-equilibrium processes in f(R) gravity. 
The coefficients of the expansion and shear terms are equal to the viscosities of the membrane paradigm\cite{Price:1986yy,Chatterjee:2010gp} 
and become those of Jacobson et al\cite{Eling:2006aw} in an infinitesimal time limit, 
and a new entropy production term appears. 

This paper is organized as follows. In section 2, Jacobson's idea and Padmanabhan's observer are introduced, 
where the difference between the two observers is explained. 
In section 3, the idea of spacetime thermodynamics developed in the section 2 is applied to a dynamical spacetime, 
and, in Einstein's gravity, the entropy balance equation of spacetime thermodynamics is derived. 
In section 4, the entropy balance equation in f(R) gravity is constructed in almost the same way. 
In section 5, conclusions and discussions are given.

In this paper, we use the units $(G=c=\hbar=k_B=1)$ and a spacetime metric with the signature $(-,+,+,+)$. 
Our sign conventions are those of MTW\cite{MTW}, with the exception of the relation between extrinsic curvature and expansion $(K^{\mu}_{\;\; \mu}=\theta)$.

%%%%%%%%%%%%%%%%%%%%%%%%%%%%%%%%%%%%%%%%%%%%%%%%%%%%%%%%%%%%%%%%%%%%%%%%%%%%%%%%%%%%%%%%%%%%%%

\section{Spacetime Thermodynamics: The Framework}
In order to consider a part of spacetime as a thermodynamic system, 
we introduce some ingredients such as observer, system, and energy flow. 
The basic idea is based on Jacobson's\cite{Jacobson:1995ab}, 
but we use Padmanabhan's observer\cite{Padmanabhan:2009vy} to measure the physics. 
\subsection{The Definition of System, External World and Heat}
In general, heat is transfer of energy which cannot be identified and controlled by an external observer. 
Therefore, in spacetime thermodynamics, heat can be defined as energy flow through any causal horizon, and this can define the system and the external world. 
That is, the system is the region inside the horizon, and the external world is the region outside the horizon. 
A conventional observer is defined as an observer in the external world, who measures thermodynamic quantities. 
He cannot identify any energy flow passed through the causal horizon, and thus, such a form of energy flow is regarded as heat for him. 
A good example is a black hole event horizon. 
An observer outside the event horizon regards the inside as the system, the outside as the external world, 
and energy flow through the horizon as heat for him. 
However, the above definition is not limited to a black hole event horizon, but applicable to any causal horizon. 
A way to construct a causal horizon at any point in any spacetime is the use of a Rindler horizon.

Rindler horizon can be constructed as follows\cite{Jacobson:1995ab,Padmanabhan:2009vy,Eling:2006aw,Chirco:2009dc}. Firstly, we take a point P in any spacetime. 
Secondly, we invoke the equivalence principle to introduce a local inertial frame for an observer near the point. 
This is always allowed if the size of the region$~l$ is restricted to $l\ll R|_P$, where $R|_P$ is the radius of curvature at P. 
The metric of this region is approximately Minkowski: 
\begin{equation}
g_{\mu \nu }=\eta_{\mu \nu}+ O(l^2)~. 
\end{equation}
Thirdly, the local patch is described by the Riemann normal coordinates$~{x^{\mu}}$, such that P stays at $x^{\mu}=0$. 
Finally, we uniformly accelerate an observer near P for the X direction. The corresponding transformation is   
\begin{equation}
T=x\sinh(\kappa t),\;\; X=x\cosh(\kappa t). 
\end{equation}
Then, the local coordinate around P is the local Rindler coordinate: 
\begin{equation}
ds^2=-\kappa ^2 x^2 dt^2+dx^2+dy^2+dz^2~,
\label{eq:LRC}
\end{equation}
where $\kappa$ is an arbitrary scaling factor. 
The 4 vector of the observer at $x=\text{const}$ is given by $u=\frac{\partial }{\partial \tau}=\frac{1}{\kappa x}\frac{\partial }{\partial t}$, 
and the proper acceleration is given by $a=\frac{1}{x}$. 
%\begin{figure}[pb]
%\centerline{\psfig{file=observer1'.eps,width=6.5cm}}
%\vspace*{8pt}
%\caption{A thermodynamic system of a spacetime and our observer\label{fig:observer1}}
%\end{figure}
\begin{figure}[h]
\begin{center}
\includegraphics[width=5.7cm,angle=0,clip]{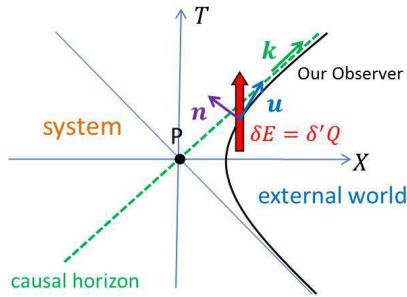}
\caption{A thermodynamic system of a spacetime and our observer}
\label{fig:observer1}
\end{center}
\end{figure}

In figure \ref{fig:observer1}, the dashed line shows the horizon that hides the inside from the view of the observer. 
Therefore, his side is the external world for him, and the other side is the system for him. 
It is the observer that measures energy flow$~\delta E$ into the system. 
Energy flow between his proper time $\tau_1$ and $\tau_2$ is given by
\begin{equation}
\delta E =\int_{\tau_1}^{\tau_2}d\tau \int_{\bar{{\cal S}}(\tau)}d\bar{A}~T_{\mu \nu}u^{\mu} n^{\nu}~,
\label{eq:energy}
\end{equation}
where $\bar{{\cal S}}(\tau)$ is the 2 dimensional spacial area of the timelike surface near the horizon at proper time$~\tau$, 
$T_{\mu \nu}$ is the energy-momentum tensor, 
$n^{\nu}$ is his normal vector inward the surface, and 
$d\bar{A}=\sqrt{\bar{h}}d\bar{x}^2$ is the area element, where $\bar{h}$ is the determinant of the spacial metric$~\bar{h}_{ab}$. 
In this paper, the symbol $\delta$ means variation in a thermodynamic quantity which occurs in the process. 
Moreover, we assume that all energy flow is heat: 
\begin{equation}
\delta E = \delta 'Q~. 
\label{eq:energy assumption}
\end{equation}
Here $\delta '$ means variation which depends upon the particular path taken through the space of thermodynamic parameters. 
Therefore, it is natural that, like heat in conventional thermodynamics, $\delta E $ given by (\ref{eq:energy}) depends on the process, that is, $u^{\mu},~n^{\mu}$ and $\bar{{\cal S}}(\tau)$. 

We here make three comments.

(a) Our observer should be sufficiently close to the point P so that 
he can take the local Rindler coordinate 
and interpret the spacetime over his horizon as the thermodynamic system. 
Therefore, we should take the observer at $x\sim 0$, and then 
energy flow$~\delta E~$ given by (\ref{eq:energy}) asymptotically becomes
\begin{equation}
\delta E =(\kappa x)^{-1}\int_{t_1}^{t_2}dt \int_{{\cal S}(t)}dA~T_{\mu \nu}k^{\mu} k^{\nu} =(\kappa x)^{-1}\delta E_{\text{K}}, 
\label{eq:energy2}
\end{equation}
where $u^{\mu}\simeq(\kappa x)^{-1}k^{\mu}$ and $n^{\mu}\simeq(\kappa x)^{-1}k^{\mu}$ for $x\sim 0$, 
$k=\frac{\partial}{\partial t}$ is horizon's generator, 
${\cal S}(t)$ is the 2 dimensional spacial area of the null horizon at time$~t$, 
and $dA=\sqrt{h}dx^2$ is the area element. 
Here $k=\frac{\partial}{\partial t}$ is the Killing vector in the Rindler coordinate \eqref{eq:LRC}, and thus, 
$\delta E_{\text{K}}$ is locally conserved energy flow through the horizon. 
In the limit where $x \rightarrow 0$, $\delta E$ diverges. 
However, the entropy balance law, which will be constructed in section 3 and 4, is finite\cite{Jacobson:1995ab}. 
Note that (\ref{eq:energy}) is integration on a timelike surface, and (\ref{eq:energy2}) is one on a null surface, 
though they are asymptotically equal in $x \sim 0$. Therefore, in the following discussion, we use $\bar{{\cal S}}(\tau) \simeq {\cal S}(t)$ and $d\bar{A} \simeq dA$ near the horizon. 

\begin{figure}%[h]
\begin{center}
\includegraphics[width=5.7cm,angle=0,clip]{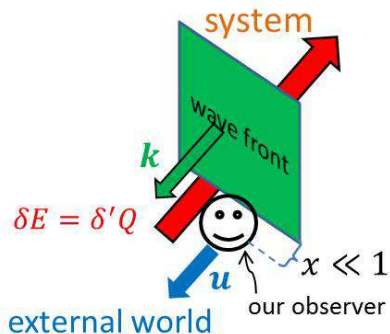}
\caption{A wave front of light and our thermodynamic system}
\label{fig:observer2}
\end{center}
\end{figure}

(b) In general, a causal horizon is a virtual wave front of light. 
Let us imagine the following situation. We take some spacial region. A virtual light emanates outward from the boundary. 
As figure \ref{fig:observer2}, our observer is accelerating in front of the wave front of the light. 
Then, he cannot observe things swallowed up by the light and can regard the energy flow as heat. 
%\begin{figure}[pb]
%\centerline{\psfig{file=observer2'.eps,width=6.5cm}}
%\vspace*{8pt}
%\caption{A wave front of light and our thermodynamic system\label{fig:observer2}}
%\end{figure}
Here we should always arrange the observer to be the same distance$~x$ from the wave front and have the constant temperature$~T_U$ in subsection 2.2, 
which means that the process can be regarded to be isothermal (see the section 3). 
Therefore, the construction of a thermodynamic system so far can also be applied to any wave front of light. 
A good example is a black hole event horizon. 
A future event horizon is defined as the boundary of the closure of the causal past of the future null infinity. 
That is, a black hole is a region from where even light cannot escape eternally, 
and the event horizon is the wave front of the light which is the boundary. 
Note that a black hole is a spatially closed thermodynamic system, but Jacobson's original system is an open system. 

(c) Our observer is the same as Padmanabhan's, not as Jacobson's. 
%\begin{figure}[pb]
%\centerline{\psfig{file=observer3'.eps,width=6.5cm}}
%\vspace*{8pt}
%\caption{The left one is Jacobson's and the right one is ours like Padmanabhan's.\label{fig:observer3}}
%\end{figure}
\begin{figure}[h]
\begin{center}
\includegraphics[width=5.7cm,angle=0,clip]{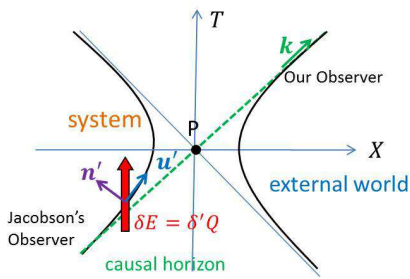}
\caption{Left one is Jacobson's and right one is ours like Padmanabhan's. }
\label{fig:observer3}
\end{center}
\end{figure}
We take Jacobson's idea, but our energy flow is not his but Padmanabhan's\cite{Padmanabhan:2009vy}. 
Jacobson measures energy flow by using ``an observer in the system" as figure \ref{fig:observer3}\cite{Jacobson:1995ab,Eling:2006aw}. 
However, it does not agree with an observer in thermodynamics. 
\footnote{Another attempt to demonstrate differences between Jacboson's and Padmanabhan's formulation is discussed in \cite{Kothawala:2010bf}.}

\subsection{Temperature and Entropy}
In this subsection, fundamental constants are introduced. 

Our observer is uniformly accelerating at $x\sim 0$ in the local Rindler coordinate. 
Then he feels the temperature of the Unruh effect\cite{Unruh:1976db}: 
\begin{equation}
T_U=\frac{\hbar a}{2\pi c k_B }=\frac{\hbar c x^{-1}}{2\pi k_B}~. 
\label{eq:unruh}
\end{equation}
He is near the system and sees it contact with the external world at the temperature. 

Next, the Rindler observer cannot get information about things which have gone into the horizon. 
This situation resembles Bekenstein's gedankenexperiment, 
who thought black hole's entropy as information defect for outside observers and derived the entropy area law\cite{Bekenstein:1973ur}. 
Thus, we assume that variation in the thermodynamic system's entropy, though observer-dependent, is Bekenstein's entropy: 
\begin{equation}
\delta S=\delta \left(\frac{k_B}{4l_p^2}\int_{{\cal S}(\tau)}dA \gamma(x)\right)~, 
\label{eq:entropy}
\end{equation}
where $l_p^2=\frac{G \hbar}{c^3}$ is the Planck area, ${\cal S}(\tau)$ is the area of the wave front of light at proper time$~\tau$, 
and $\gamma(x)$ is the entropy density on it. 
For example, in the case of Einstein's gravity, $\gamma(x)=1$.

Therefore, instantaneous equilibrium condition is given by 
\begin{equation}
\left.\frac{dS}{d\tau}\right |_{\tau}=0~.
\label{eq:condition}
\end{equation}

%%%%%%%%%%%%%%%%%%%%%%%%%%%%%%%%%%%%%%%%%%%%%%%%%%%%%%%%%%%%%%%%%%%%%%%%%%%%%%%%%%%%%%%%%%%%%%%%%%%%%%%%%

\section{Entropy Balance Equation of Spacetime Thermodynamics in Einstein's gravity}
Let's derive the entropy balance equation of spacetime thermodynamics for non-equilibrium processes in Einstein's gravity. 
We use, as Jacobson, the Raychaudhuri equation, not as him, an outside observer near the causal horizon and his proper time. 
We will not derive the Einstein equation but construct the entropy balence equation by using the Einstein equation. 
This approach is based on the conventional derivation of black hole's first law for quasi-static processes\cite{Poisson,Carter}. 

\subsection{Derivation}
We first estimate the change of the Rindler horizon's area. 
The Raychaudhuri equation for null congruence in a non-affine parameter$~t$ is\cite{Poisson}:
\begin{equation}
\frac{d \theta}{dt}=\kappa \theta - \frac{\theta ^2}{2}-\sigma_{\mu \nu}\sigma^{\mu \nu}-R_{\mu \nu}k^{\mu}k^{\nu}~,
\label{eq:ray1}
\end{equation}
where $k=\frac{\partial}{\partial t}$ is null generator of the horizon, $\theta$ is the expansion, 
$\sigma _{\mu \nu}$ is the shear tensor, and $\kappa$ is defined as $k^{\mu}_{\;\; ;\nu}k^{\nu}=\kappa k^{\mu}$. 
Note that we here use the Raychaudhuri equation not for timelike congruence but for null congruence, 
because we will formulate the entropy balance equation of the spacetime thermodynamic system constructed by the null horizon.
%However, in this spacetime, unlike a static black hole spacetime, the global Killing vector does not exist, 
%so we use the 4 vector$~u^{\mu}$ of the observer near the local horizon to estimate energy flow$~\delta E$ as follows. 
Expansion$~\theta$ can also be written as\cite{Poisson}
\begin{equation}
\theta = \frac{1}{\Delta A}\frac{d \Delta A}{dt}~,
\end{equation}
where $\Delta A=\sqrt{h}(\Delta x)^2$ is the area element of the wave front, and $h$ is the determinant of the spacial metric on it. 
By this, (\ref{eq:ray1}) is expressed as
\begin{equation}
\frac{d^2 \Delta A}{dt^2}=\left(\kappa \theta +\frac{\theta ^2}{2}-\sigma_{\mu \nu}\sigma^{\mu \nu}-R_{\mu \nu}k^{\mu}k^{\nu} \right)\Delta A~.
\label{eq:ray2}
\end{equation}
Here we transform from $t$ into the observer's proper time$~\tau$: 
\begin{equation}
\frac{d^2 \Delta A}{d\tau^2}=\left(x^{-1} \bar{\theta} +\frac{\bar{\theta} ^2}{2}-\bar{\sigma}_{\mu \nu}\bar{\sigma}^{\mu \nu}-R_{\mu \nu}\bar{k}^{\mu}\bar{k}^{\nu} \right)\Delta A~,
\label{eq:ray3}
\end{equation}
where $d\tau =\kappa x dt,\; \theta=\frac{\partial \tau}{\partial t}\bar{\theta}=\kappa x \bar{\theta},\; \sigma_{\mu\nu}=\frac{\partial \tau}{\partial t}\bar{\sigma}_{\mu\nu}=\kappa x \bar{\sigma}_{\mu\nu},$ and $k^{\mu}=\kappa x\bar{k}^{\mu}$. 
Note that as discussed in the comment (b) of the subsection 2.1, 
we always arrange the observer to be the same distance$~x$ from the horizon, and then, we can regard $x^{-1}$ as a constant. 
Then we perform area integral on the horizon at $\tau$: 
\begin{equation}
\frac{d^2 A}{d\tau^2}=x^{-1} \frac{dA}{d\tau} +\int_{{\cal S}(\tau)}dA\left(\frac{\bar{\theta} ^2}{2}-\bar{\sigma}_{\mu \nu}\bar{\sigma}^{\mu \nu}-R_{\mu \nu}\bar{k}^{\mu}\bar{k}^{\nu} \right)~,
\label{eq:ray4}
\end{equation}
where $A=\int_{{\cal S}(\tau)}dA$. Then, we use the time-independence of $x^{-1}$ and perform time integral between $\tau_{1}$ and $\tau_{2}$:
\begin{equation}
\left.\frac{dA}{d\tau}\right|^{\tau_2}_{\tau_1}=x^{-1} \left.A\right|^{\tau_2}_{\tau_1} +\int_{\tau_1}^{\tau_2}d\tau \int_{{\cal S}(\tau)}dA\left(\frac{\bar{\theta} ^2}{2}-\bar{\sigma}_{\mu \nu}\bar{\sigma}^{\mu \nu}-R_{\mu \nu}\bar{k}^{\mu}\bar{k}^{\nu} \right)~.
\label{eq:ray5}
\end{equation}
The Einstein equation
\begin{equation}
R_{\mu\nu}-\frac{1}{2}Rg_{\mu\nu}=8\pi T_{\mu\nu}
\end{equation}
and the null vector $k$ lead to 
\begin{eqnarray}
\frac{x^{-1}}{2\pi}\frac{1}{4}\delta A =\frac{1}{8\pi}\delta \left(\frac{dA}{d\tau}\right)+\frac{1}{8\pi}\int_{\tau_1}^{\tau_2}d\tau \int_{{\cal S}(\tau)}dA\left(-\frac{\bar{\theta} ^2}{2}+\bar{\sigma}_{\mu \nu}\bar{\sigma}^{\mu \nu}\right)\nonumber\\
+(\kappa x)^{-1}\int_{t_1}^{t_2}dt \int_{{\cal S}(t)}dA~T_{\mu \nu}k^{\mu}k^{\nu}~.
\label{eq:ray6}
\end{eqnarray}
Finally, we use the asymptotic expression (\ref{eq:energy2}) of energy flow $\delta E$, 
the entropy formula (\ref{eq:entropy}) for $\gamma=1$ and the Unruh temperature (\ref{eq:unruh}):  
\begin{equation}
T_U\delta S =\frac{1}{2\pi}\delta \left(\frac{dS}{d\tau}\right)+\frac{1}{8\pi}\int_{\tau_1}^{\tau_2}d\tau \int_{{\cal S}(\tau)}dA\left(-\frac{\bar{\theta} ^2}{2}+\bar{\sigma}_{\mu \nu}\bar{\sigma}^{\mu \nu}\right)+\delta E~.
\label{eq:ray7}
\end{equation}
Here we assume that the initial and final state are equilibrium, and then, the instantaneous equilibrium condition (\ref{eq:condition}) leads to
\begin{equation}
\delta S =\frac{1}{T_U}\int_{\tau_1}^{\tau_2}d\tau \int_{{\cal S}(\tau)}dA\left(\frac{-1}{16\pi}\bar{\theta} ^2+\frac{1}{16\pi}2\bar{\sigma}_{\mu \nu}\bar{\sigma}^{\mu \nu}\right)+\frac{\delta 'Q}{T_U} ,
\label{eq:formula1}
\end{equation}
where we use the assumption \eqref{eq:energy assumption}. 
This is the entropy balance equation of spacetime thermodynamics for non-equilibrium processes in Einstein's gravity.

\subsection{Interpretation}
Before considering  the meaning of (\ref{eq:formula1}), we review the entropy balance equation.
The second law is, in Clausius's form, 
\begin{equation}
T^{(ex)}\delta S \geq \delta 'Q~,
\label{eq:2nd}
\end{equation}
where $\delta S$ is the variation in system's entropy, $T^{(ex)}$ is the external temperature which is constant in the process, 
$\delta 'Q$ is the heat from the external world to the system. 
If the process is quasi-static, $T^{(ex)}=T$ and $T\delta S=\delta 'Q$, where $T$ is the temperature of the system. 

This can also be written as the entropy balance equation: 
\begin{equation}
\delta S = \frac{\delta 'Q}{T^{(ex)}} + \delta 'D,\;\; \delta 'D \geq 0~,
\label{eq:2nd2}
\end{equation}
where $\delta 'D$ is internal entropy production, such as friction, diffusion, and heat conduction\cite{Groot}. 
If the process is quasi-static process, $T^{(ex)}=T$ and $\delta 'D=0$. 

Let's compare the formula (\ref{eq:formula1}) with the basic equation (\ref{eq:2nd2}). 
Firstly, it is clear that $\delta'Q$ is considered as the heat and 
$\delta S$ as the variation of th entropy. 
Secondly, we assume that our observer near the horizon feels the Unruh temperature $T_U$, which is constant as mentioned in the comment (b) of subsection 2.1. 
Thus we can regard the temperature as the external temperature $T^{(ex)}$ at which the system contacts with the external world, 
which means that the observer can regard the process as isothermal. 
Thirdly, the shear$~\bar{\sigma}_{\mu \nu}$ comes mainly from Weyl tensor in not-too-dynamical processes\cite{Poisson}, 
which is pure gravitational degrees of freedom. 
%So, the shear term$~2\bar{\sigma}_{\mu \nu}\bar{\sigma}^{\mu \nu}$ would be a reversible mechanical term 
%and would correspond to the work$~\delta 'W$. 
%However, in the system whose thermodynamic parameter is energy, $\delta 'W = \delta 'D \geq 0$, as mentioned below (\ref{eq:2nd2}). 
The term is always positive ($\bar{\sigma}_{\mu \nu}\bar{\sigma}^{\mu \nu}\geq 0$) and increases the entropy for any dynamical process. 
This corresponds to the fact that entropy production from the squared gradients of state variables is a universal property of non-equilibrium thermodynamics\cite{Groot}. 
Moreover, in the case of black hole, this term coincides with the Hartle-Hawking formula for the tidal heating of a classical black hole\cite{Chirco:2009dc,Poisson,Hawking:1972hy}. 
Thus, the shear term corresponds to $\delta 'D$. 
Finally, from (\ref{eq:entropy}) for $\gamma =1$, the expansion $\bar{\theta}$ is the density of entropy increase per unit proper time, 
so the term can regarded as an entropy production term. 
Therefore, both of them can be included in $\delta 'D$. 

There are some remarks. 

(a)The expansion term's coefficient$~\zeta$ and the shear term's one$~\eta$ are respectively
\begin{equation}
\zeta = -\frac{1}{16\pi},\;\; \eta =\frac{1}{16\pi}~,
\label{eq:viscosity}
\end{equation}
which are equal to viscosities of the black hole membrane paradigm\cite{Price:1986yy}. 
Indeed, our observer corresponds to the FIDO in the membrane paradigm applied to the Rindler horizon. 
Therefore, our observer can regard the timelike surface$~\bar{{\cal S}}(\tau)$ as some viscous fluid. 

(b)In the limit $x\rightarrow 0$, the temperature ($T_U=\frac{x^{-1}}{2\pi}$) and the heat ($\delta 'Q \propto x^{-1}$) diverge, 
and then, (\ref{eq:formula1}) seems to be singular. 
However, this is found to be finite if we use $d\tau = \kappa xdt$ and rewritten it in terms of the Killing time $t$ as
\begin{equation}
\delta S =\frac{1}{T_K}\int_{t_1}^{t_2}dt \int_{{\cal S}(t)}dA\left(\frac{-1}{16\pi}{\theta} ^2+\frac{1}{16\pi}2{\sigma}_{\mu \nu}{\sigma}^{\mu \nu}\right)+\frac{\delta 'Q_K}{T_K} ,
\label{eq:formula1'}
\end{equation}
where $T_K=\frac{\kappa}{2\pi}$ and $\delta'Q_K=\delta E_K$. 
Note that $\kappa$ is an arbitrary scaling factor and $t$ is not uniquely determined, 
but \eqref{eq:formula1'} is invariant under the following scaling:
\begin{equation}
t\rightarrow \alpha t \; \text{ and }\; \kappa \rightarrow \alpha^{-1}\kappa.
\end{equation}
Therefore, \eqref{eq:formula1'} does not depend on the choice of $t$ and $\kappa$. 
Note that this result comes from the fact that $d\tau = \kappa xdt$ is invariant. 

(c) We estimate all quantities in proper time$~\tau$ of the observer near the horizon. 
Why do we use not $t$ but rather $\tau$ ? 
In the local spacetime thermodynamic system for non-equilibrium processes, unlike in a black hole system for quasi-static proceses, 
the global Killing vector does not exist, and thus, 
the local Killing time $t$ and the locally conserved energy flow $\delta E_K$ cannot be normalized uniquely. 
On the other hand, the proper time $\tau$ is always normalized as $u^2=-1$, and 
$\delta E$ is uniquely determined. 
Therefore, the use of $\tau$ clarifies the physical meaning of the thermodynamic system, which is  observer-dependent.

%(a) The derivation of (\ref{eq:formula1}) is applicable to any spacetime thermodynamic system in section 2. 
%For the initial and final state, the equilibrium condition (\ref{eq:condition}) is assumed there. 
%Locally, this is equal to $\bar{\theta}|_{\tau_1, \tau_2} = 0$. 
%If we take such a small ``patch" system of a black hole event horizon that it can be interpreted as a local inertial frame, 
%this condition is automatically satisfied, 
%for the event horizon of a stationary black hole coincides the apparent horizon, which has $\bar{\theta}=0$. 
%Therefore, (\ref{eq:formula1}) for the above system has the most physical meaning.
%However, if the assumption (\ref{eq:condition}) is made, (\ref{eq:formula1}) can be, in principle, 
%applied to any spacetime thermodynamic system, though it may be an open system. 

(d) In the case of black hole thermodynamics, 
an observer at infinity considers the origin of any change of the entorpy as the chage of the ADM energy\cite{Poisson}. 
In dynamical spacetime, it is difficult to distinguish matter energy and gravitational energy from each other. 
On the other hand, our formula \eqref{eq:formula1} is exact near the horizon as long as the system can be described in the Rindler coordinate. 
Therefore our observers near the horizon can distinguish which the change of the entropy comes from matter flow or purely gravitational one. 
That is because $T_{\mu \nu}$ in $\delta'Q$ corresponds to matter energy flow into/from the horizon, and 
$\bar{\theta}^2$ and $\bar{\sigma}_{\mu \nu}^2$ correspond to the horizon dynamics, which produces purely gravitational entropy production.

%(c)Our formula is exact, so it can be applied to any dynamical processes and contains both matter and gravitational energy flow, 
%which, if the system is a small patch system of a black hole event horizon in (a), contribute to the variation$~\delta M_{BH}$ of the ADM mass of the black hole. 
%Indeed, in quasi-static processes, (\ref{eq:formula1}) becomes 
%\begin{equation}
%T_U \delta S = \delta E~,
%\label{eq:quasi}
%\end{equation}
%where, if the variation is so small that the Hawking temperature$~T_{BH}$ can be regarded as constant in the process, 
%$T_U=\frac{T_{BH}^{(initial)}}{\sqrt{g_{00}^{(initial)}(r\sim r_{H})}}$. 
%This is the blue-shifted conventional first law of black hole thermodynamics, 
%in which only matter energy flow are taken into account through $T_{\mu\nu}$. 
%Note that the Hartle-Hawking formula contains only the leading gravitational energy in not-too-dynamical processes \cite{Hawking:1972hy}. 

(e) The expansion term's coefficient$~\zeta$ is negative. 
This is one of properties of the Raychaudhuri equation (\ref{eq:ray1}) for null congruence in a non-affine parameter\cite{Carter}. 
The negative coefficient would imply that the entropy of the system can decease and the system can become thermodynamically unstable. 
However, this does not occur at least classically. 
The first reason is that the equation is teleological 
in the sense that it should be subject not to the initial condition $\bar{\theta}(0)=0$ but to the final condition $\bar{\theta}(\infty)=0$\cite{Carter}. 
In the case of a black hole, the teleological property is natural 
because the event horizon is globally defined, and so, determining its location at a time requires all the future information\cite{Price:1986yy,Carter}. 
Note that our basic time scale is $x\ll 1$, and so, if the spacetime thermodynamic system is instantaneously equilibrium at $\tau_1$ and $\tau_2$, 
the teleological effect can be neglected by using the discussion by Carter\cite{Carter}. 
The second reason, which is applicable only to black holes, is that, from Hawking's area theorem\cite{Hawking:1971tu}, 
the black hole area never decreases classically. 
In a stationary black hole the event horizon is the same as the apparent horizon, and thus, 
in a spherical process ($\bar{\sigma}_{\mu\nu}=0$), $\bar{\theta}\neq 0$ must be accompanied by $\delta E \neq 0$. 
Therefore, no matter how dynamical the spherical process is, 
$\delta E$ must be larger than the absolute value of the expansion term in (\ref{eq:formula1}), in order to increase the area. 
Thus a black hole is thermodynamically stable in processes without the Hawking radiation. 
The negative coefficient in quantum effects will be discussed in section 5. 

(f) If the logic of our derivation is reversed, 
the Einstein equation can be derived as ``the equation of state" from the entropy balance equation (\ref{eq:formula1}) 
for non-equilibrium processes, as Jacobson\cite{Jacobson:1995ab,Eling:2006aw,Chirco:2009dc}. 
However, we use the outside observer consistently, unlike him, and thus, the physical meaning is more clear. 

%%%%%%%%%%%%%%%%%%%%%%%%%%%%%%%%%%%%%%%%%%%%%%%%%%%%%%%%%%%%%%%%%%%%%%%%%%%%%%%%%%%%%%%%%%%%%%%%%

\section{Entropy Balance Equation of Spacetime Thermodynamics in f(R) Gravity}
Now we consider to the entropy balance equation in f(R) gravity. 
\subsection{Derivation}
The physical situation and the derivation are almost the same as the previous section. 
The main differences are ``the equation of state" and the entropy density. 
f(R) gravity is defined by the action\cite{Sotiriou:2008rp}: 
\begin{equation}
I=\frac{1}{16\pi}\int d^4x \sqrt{-g}f(R)~,
\label{eq:action}
\end{equation}
where $f(R)$ is an arbitrary function of Ricci scalar $R$. Variation principle gives the equation of motion:
\begin{equation}
f'(R)R_{\mu\nu}-\nabla _{\mu} \nabla _{\nu} f'(R)+\left(\nabla ^2 f'(R)-\frac{1}{2} f(R) \right) g_{\mu\nu}=8\pi T_{\mu\nu}~,
\label{eq:eom}
\end{equation}
where $f'(R)\equiv \frac{df}{dR}$. The black hole's entropy is given by the Wald entropy\cite{Wald:1993nt}:
\begin{equation}
S=\frac{1}{4}\int_{{\cal S}} dA~f'(R)~.
\label{eq:entropy-f1}
\end{equation}
The variance is
\begin{eqnarray}
\delta S&=&\frac{1}{4}\int_{t_1}^{t_2}dt \left(\int_{{\cal S}(t)}\frac{dA}{dt}f'(R)+\int_{{\cal S}(t)}dA\frac{df'(R)}{dt} \right)\nonumber\\
&=&\frac{1}{4}\int_{t_1}^{t_2}dt \int_{{\cal S}(t)}dA\left(\theta f'(R)+\frac{df'(R)}{dt} \right)~,
\label{eq:entropy-f2}
\end{eqnarray}
which is (\ref{eq:entropy}) for $\gamma = f'(R)$. 

Let's derive the entropy balance equation. 
First, multiplying (\ref{eq:ray3}) by $f'(R)$, we obtain 
\begin{equation}
f'\frac{d^2 \Delta A}{d\tau^2}=x^{-1} f'\bar{\theta} \Delta A +f'\left(\frac{\bar{\theta} ^2}{2}-\bar{\sigma}_{\mu \nu}\bar{\sigma}^{\mu \nu}\right)\Delta A -f'R_{\mu \nu}\bar{k}^{\mu}\bar{k}^{\nu} \Delta A~.
\label{eq:ray'2}
\end{equation}
Using (\ref{eq:eom}) and (\ref{eq:entropy-f2}), we rewrite (\ref{eq:ray'2}) as 
\begin{eqnarray}
f'\frac{d^2 \Delta A}{d\tau^2}=4x^{-1} \frac{1}{4} \left(f'\bar{\theta} + \frac{df'}{d\tau}\right)\Delta A -x^{-1} \frac{df'}{d\tau} \Delta A+
f'\left(\frac{\bar{\theta} ^2}{2}-\bar{\sigma}_{\mu \nu}\bar{\sigma}^{\mu \nu}\right)\Delta A \nonumber\\
-\bar{k}^{\mu}\bar{k}^{\nu}\nabla _{\mu} \nabla _{\nu} f'\Delta A-8\pi T_{\mu\nu}\bar{k}^{\mu}\bar{k}^{\nu}\Delta A~,
\label{eq:ray'3}
\end{eqnarray}
where we use $\bar{k}^2 =0$. After the area integral, we perform the time integral, and thus obtain
\begin{eqnarray}
4x^{-1} \delta S+\int_{\tau_1}^{\tau_2}d\tau \int_{{\cal S}(\tau)}dA~f'\left(\frac{\bar{\theta} ^2}{2}-\bar{\sigma}_{\mu \nu}\bar{\sigma}^{\mu \nu}\right)
-8\pi \int_{\tau_1}^{\tau_2}d\tau\int_{{\cal S}(\tau)}dA~T_{\mu\nu}\bar{k}^{\mu}\bar{k}^{\nu}\nonumber\\
=\int_{\tau_1}^{\tau_2}dt\int_{{\cal S}(\tau)}\frac{d^2A}{d\tau^2}f'+\int_{\tau_1}^{\tau_2}d\tau \int_{{\cal S}(\tau)}dA \left(x^{-1} \frac{df'}{d\tau} +\bar{k}^{\mu}\bar{k}^{\nu}\nabla _{\mu} \nabla _{\nu} f'\right)~.
\label{eq:ray'4}
\end{eqnarray}
Now we have
\begin{equation}
\frac{df'(R)}{d\tau}=f''(R)R_{;\mu}u^{\mu}\simeq f''(R)R_{;\mu}\bar{k}^{\mu}
\end{equation}
and $\bar{k}^{\mu}_{\;\; ;\nu}\bar{k}^{\nu}=x^{-1} \bar{k}^{\mu}$, and thus, reach
\begin{equation}
\frac{d^2f'}{d\tau^2}=\bar{k}^{\mu}\bar{k}^{\nu}\nabla _{\mu} \nabla _{\nu} f'+x^{-1} \frac{df'}{d\tau}~.
\end{equation}
From this and (\ref{eq:entropy-f2}),
\begin{eqnarray}
 \verb|RHS of |(\ref{eq:ray'4})&=&\int_{\tau_1}^{\tau_2}d\tau \int_{{\cal S}(\tau)}\frac{d^2A}{d\tau^2}f'+\int_{\tau_1}^{\tau_2}d\tau \int_{{\cal S}(\tau)}dA \frac{d^2f'}{d\tau^2}\nonumber\\
 &=&  \int_{\tau_1}^{\tau_2}d\tau\frac{d}{d\tau}\int_{{\cal S}(\tau)}\left(\frac{dA}{d\tau}f'+dA \frac{df'}{d\tau}\right)-2\int_{\tau_1}^{\tau_2}d\tau\int_{{\cal S}(\tau)}\frac{dA}{d\tau} \frac{df'}{d\tau}\nonumber \\
 &=&  4\delta \left( \frac{dS}{d\tau}\right)-2\int_{\tau_1}^{\tau_2}d\tau \int_{{\cal S}(\tau)}dA~\bar{\theta} \frac{df'}{d\tau}~.
\end{eqnarray}
Using this, $u\simeq n \simeq \bar{k}$, and (\ref{eq:energy2}), we rewrite (\ref{eq:ray'4}) as follows: 
\begin{eqnarray}
\frac{x^{-1}}{2\pi}\delta S =\frac{1}{2\pi}\delta \left(\frac{dS}{d\tau}\right)-\frac{1}{4\pi}\int_{\tau_1}^{\tau_2}d\tau \int_{{\cal S}(\tau)}dA~\bar{\theta} \frac{df'}{d\tau}\nonumber\\
+\frac{1}{16\pi}\int_{\tau_1}^{\tau_2}d\tau \int_{{\cal S}(\tau)}dA~f'(-\bar{\theta} ^2+2\bar{\sigma}_{\mu \nu}\bar{\sigma}^{\mu \nu})+\delta E~.
\label{eq:ray'5}
\end{eqnarray}
By \eqref{eq:energy assumption}, (\ref{eq:unruh}) and the assumption that the initial and final state are equilibrium, 
we finally arrive at 
\begin{eqnarray}
\delta S =\frac{1}{T_U}\int_{\tau_1}^{\tau_2}d\tau \int_{{\cal S}(\tau)}dA\left(\frac{-f'(R)}{16\pi}\bar{\theta} ^2+\frac{f'(R)}{16\pi}2\bar{\sigma}_{\mu \nu}\bar{\sigma}^{\mu \nu}\right)\nonumber\\
-\frac{1}{4\pi T_U}\int_{\tau_1}^{\tau_2}d\tau \int_{{\cal S}(\tau)}dA~\bar{\theta} \frac{df'(R)}{d\tau}+\frac{\delta'Q}{T_U}~.
\label{eq:formula2}
\end{eqnarray}
This is the entropy balance equation generalized to f(R) gravity. 

\subsection{Interpretation}
The meaning of (\ref{eq:formula2}) is essentially the same as the case of Einstein's gravity. 
In (\ref{eq:formula2}), the expansion term's coefficient$~\zeta$ and the shear term's one$~\eta$ are found as 
\begin{equation}
\zeta = -\frac{1}{16\pi}f'(R),\;\; \eta =\frac{1}{16\pi}f'(R)~,
\label{eq:viscosity2}
\end{equation}
which are equal to the viscosities of the black hole membrane paradigm in f(R) gravity\cite{Chatterjee:2010gp}. 
%in which the first law (\ref{eq:formula2}) was not derived and only a momentum conservation equation like the Navier-Stokes equation was obtained. 
%In this sense, the equivalence in subsection 3.2 between our observer's viewpoint and the black hole membrane paradigm holds in f(R) gravity, too. 

There are some remarks.

(a) $\zeta(x)$ and $\eta(x)$ depend on the spacetime point$~x$, which comes from the entropy density $\gamma(x)=f'(R(x))$. 
The spacetime dependence of the entropy density and viscosities may reflect a microscopic structure of spacetime 
because f(R) gravity includes higher-curvature terms and 
their coefficients are determined by renormalization of quantum field in the curved spacetime\cite{QFT}. 

(b) The following term inevitably arises in (\ref{eq:formula2}):
\begin{equation}
-\frac{1}{4\pi T_U}\int_{\tau_1}^{\tau_2}d\tau \int_{{\cal S}(\tau)}dA \bar{\theta} \frac{df'(R)}{d\tau}~.
\label{eq:newterm}
\end{equation}
This term does not depend on $T_{\mu\nu}$. 
It is second derivatives respective with proper time$~\tau$ of our observer, 
and thus, if quasi-static process, it vanishes as fast as the expansion term$~\bar{\theta}^2$ and the shear term$~\bar{\sigma}_{\mu\nu}^2$. 
In this sense, this term is effective only in non-equilibrium processes. 
In contrast to $\bar{\theta}^2$ and $\bar{\sigma}_{\mu\nu}^2$, 
the sign is not fixed and can be both positive and negative, depending on the process. 
%Now, this spacetime thermodynamic system has a single thermodynamic parameter, that is, energy. 
Therefore, we can conjecture that the term corresponds to a new internal entropy production term of $\delta 'D$ in (\ref{eq:2nd2}). 
However, the meaning is not clear yet, which will be discussed in the next section. 

(c) Our bulk viscosity$~\zeta$ in (\ref{eq:viscosity2}) is equal to that of the membrane paradigm, 
but not to that of Jacobson et al\cite{Eling:2006aw}, which is $\zeta =\frac{3}{16\pi}f'(R)$. 
They used $\lambda _0$ such that $\left. \frac{dS}{d\lambda}\right |_{\lambda_0}=0$, and expanded the equation around it, 
where $\lambda$ was the affine parameter. 
Therefore, in order to reproduce the same situation in our formula (\ref{eq:formula2}), 
we take $\tau_1=\tau_0,\; \tau_2=\tau_0+\delta \tau,\; \delta \tau \ll 1$ and expand our formula. 
Note that $\delta \lambda \propto \delta \tau $, so this limit corresponds to the same situation. 
Then we use
\begin{equation}
\bar{\theta}f'+\frac{df'}{d\tau}=0\;\;\verb|for|\;\; \tau=\tau_0
\end{equation}
and eliminate $\frac{df'}{d\tau}$ in (\ref{eq:formula2}). We arrive at 
\begin{equation}
\delta S =\frac{1}{T_U}\int_{\tau_0}^{\tau_0+\delta \tau}d\tau \int_{S(\tau)}dA\left(\frac{3f'(R)}{16\pi}\bar{\theta} ^2+\frac{f'(R)}{16\pi}2\bar{\sigma}_{\mu \nu}\bar{\sigma}^{\mu \nu}\right)+\frac{\delta'Q}{T_U}~.
\label{eq:formula2'}
\end{equation}
Thus, we obtain
\begin{equation}
 \zeta =\frac{3f'(R)}{16\pi}~,
\end{equation}
which is the same as that of Jacobson et al. 
This meaning will be discussed in the next section. 

%%%%%%%%%%%%%%%%%%%%%%%%%%%%%%%%%%%%%%%%%%%%%%%%%%%%%%%%%%%%%%%%%%%%%%%%%%%%%%%%%%%%%%%%

\section{Conclusions and Discussions}
We have applied Padmanabhan's observer to a dynamical spacetime, used the Raychaudhuri equation in a non-affine parameter, and 
generalized the formulation of spacetime thermodynamic system to the extent that we can consider non-equilibrium processes from viewpoint of the outside observer. 
Using this formulation, we have constructed (\ref{eq:formula1}) and (\ref{eq:formula2}), that is, the entropy balance equation of spacetime thermodynamics 
for non-equilibrium processes in Einstein's gravity and f(R) gravity, respectively. 
They are exact near the horizon as long as the system can be described in the Rindler coordinate, 
and take into accout both matter and gravitational energy.
The coefficients of the expansion and shear terms are equal to the viscosities of the black hole membrane paradigm\cite{Price:1986yy,Chatterjee:2010gp}, 
and a new term (\ref{eq:newterm}) appears, which is conjectured as a new internal entropy production term in $\delta'D$. 
Moreover, in the infinitesimal time limit, our coefficients agree with those of Jacobson et al\cite{Eling:2006aw}. 

There remain some open questions. 

(a) We should understand the expansion$~\bar{\theta}$ in f(R) gravity more correctly. 
Though in Einstein's gravity $\bar{\theta}$ corresponds to the density of entropy increase per unit proper time, 
in f(R) gravity this interpretation is not correct. 
Unlike the shear$~\bar{\sigma}_{\mu \nu}$ which is interpreted as purely gravitational effect, 
the expansion$~\bar{\theta}$ is sensitive to $T_{\mu \nu}$, $\bar{\theta}$ and $\bar{\sigma}_{\mu \nu}$, so the meaning is less clear. 
The remarks (b) and (c) in the previous section are intimately related to the above fact. 
$\zeta(x)$ in (\ref{eq:viscosity2}) is equal to the bulk viscosity in the membrane paradigm\cite{Chatterjee:2010gp}, 
and the new term in (\ref{eq:newterm}) inevitably arises in (\ref{eq:formula2}). 
Note that even if we rewrite the new term as $4\int_{\tau_1}^{\tau_2}d\tau \int_{{\cal S}(\tau)}dA \bar{\theta} \frac{d\zeta(x(\tau))}{d\tau}$, 
this is not correct because the spacetime-dependence of the bulk viscosity is already contained in (\ref{eq:formula2}), as ordinary fluid\cite{Landau}. 
In the infinitesimal time limit, the new term disappears and $\zeta(x)$ becomes that of Jacobson et al. 
In these senses, the interpretations of the expansion term$~\bar{\theta}^2$ and the new term may depend on the time scale of the process. 

(b) In black hole thermodynamics, the generalized second law plays a fundamental role, 
which is an assumption made by Bekenstein\cite{Bekenstein:1973ur} that 
the sum of the black hole entropy $S_{BH}=\frac{1}{4}A$ and the entropy $S_{matter}$ of the usual matter and gravitational radiation 
outside a black hole never decreases. 
Though an explicit general proof of this law has not been given until now, the validity of the law for special cases have been verified, 
such as quasi-static processes without the back-reaction of quantum field energy taken into account\cite{Frolov:1993fy}. 
In a full proof, arbitrary dynamical processes and the back-reaction should be considered.  
Therefore, our formulae may be useful to prove the generalized second law for dynamical processes 
because they can be applied to any dynamical processes with the Hawking radiation, 
though they are applicable only to a small system near the horizon. 
Especially, when the effect of the evaporation is large in the process such as the evaporation process of a small black hole, 
the expansion term$~\bar{\theta}^2$ can become more effective. 

(c) What is the entropy balance equation for more general theory of gravity? 
Our formula is restricted to f(R) gravity, which is the simplest model in higher-curvature theories of gravity. 
However, a complete proof of the generalized second law should need more general higher-curvature terms, 
such as the Gauss-Bonnet term, due to back reaction from quantum field renormalization\cite{QFT}. 
Thus, we are interested in the entropy balance equation for non-equilibrium processes in the Lovelock gravity\cite{Lovelock:1971yv}, 
which is the most general second-order gravity theory in higher dimensional spacetime. 
Note that the quasi-static processes has been studied through the Wald entropy\cite{Padmanabhan:2009vy}. 

These issues require further study. 

%%%%%%%%%%%%%%%%%%%%%%%%%%%%%%%%%%%%%%%%%%%%%%%%%%%%%%%%%%%%%%%%%%%%%%

\section*{Acknowledgments}
The author is grateful to H.~Kawai for valuable discussions. 
The author also thanks T.~Shiromizu, K.~Takae, and S.~Toh for helpful comments. 
This work is supported by the Grant-in-Aid for the Global COE 
program ``The Next Generation of Physics, Spun from Universality and Emergence" from the MEXT, 
and the Japan Society for the Promotion of Science (JSPS).

%\begin{thebibliography}{000} %for 3 digits
%\begin{thebibliography}{00}  %for 2 digits

\end{document}